\documentclass{elsart}
\usepackage{graphicx}
\usepackage{amssymb}
\newcommand{\be}{\begin{equation}}
\newcommand{\ee}{\end{equation}}
\newcommand{\bea}{\begin{eqnarray}}
\newcommand{\eea}{\end{eqnarray}}

\newcommand{\w}{\omega}
\newcommand{\es}{\epsilon_s}
\newcommand{\eu}{\epsilon_u}

\begin{document}
\begin{frontmatter}
\title{Does a surface attached globule phase exist ?} 
\author{P. K. Mishra$^1$, D. Giri$^2$, S. Kumar$^1$ 
and Y. Singh$^1$}  
\address{{\small $^1$ Department of Physics, Banaras Hindu University,
Varanasi 221 005, India } \\
{\small $^2$ Centre for Theoretical Studies, I. I. T., Kharaghpur-721302,
W. B., India}}
\begin{abstract}
A long flexible neutral polymer chain immersed in a poor solvent
and interacting with an impenetrable attractive surface exhibits a 
phase known as surface attached globule ({\bf SAG}) in addition to 
other adsorbed and desorbed phases. In the thermodynamic limit, the 
{\bf SAG} phase has the same free energy per monomer as the globular 
phase, and the transition between them is a surface transition. We 
have investigated the phase diagrams of such a chain in both two- 
and three- dimensions and calculated the distribution of monomers
in different domains of the phase diagram. 
\end{abstract}                                                               

\begin{keyword}
Surface attached globule({\bf SAG}) phase; Exact enumeration.
\PACS 64.60.-i,68.35.Rh,5.50.+q 
\end{keyword}
\end{frontmatter}


A long flexible neutral polymer chain immersed in a poor solvent and 
interacting with an impenetrable surface is known to exhibit a very rich 
phase diagram \cite{1}. Competition between the lower internal energy 
near an attractive wall and higher entropy away from it results in a 
transition, where for a strongly attractive surface the polymer sticks 
to the surface and for weak attraction the polymer chain remains desorbed. 
On the other hand, due to the self-attraction in the polymers the possibility 
of a collapse transition both in the desorbed and adsorbed states occur. 
The phase diagram is generally plotted in variables $\omega= e^{\beta 
\epsilon_s}$ and $u= e^{\beta \epsilon_u}$ where $\beta$ is the inverse 
temperature, $\epsilon_s$ is the (attractive) energy associated with each 
monomer lying on the surface and $\epsilon_u$ represents an attractive 
interaction energy between pairs of monomer which come close to each other 
and are separated along the chain by more than one unit. The short range 
repulsion between the monomer is taken in a lattice model by self-avoidance
\cite{2}. Considerable attention has been paid to find the full phase diagram 
of such a chain in both two dimensions ($2D$) and three dimensions 
($3D$).

In one of the earliest papers on the subject, Bouchaud and Vannimenus \cite{3} 
derived the exact phase diagram on a $3D$-Sierpinski gasket. The phase diagram 
consisted of the adsorbed expanded ({\bf AE}), desorbed expanded ({\bf DE}) 
and desorbed collapsed ({\bf DC}) phases. Kumar and Singh 
\cite{4} have reinvestigated 
the  phase diagram for $3D$-Sierpinski gasket and showed that for a certain 
range of surface interactions, an additional  phase having the 
feature of globule attached to 
the surface exists. Recently Singh et.al. \cite{5,6} using extrapolation of 
exact series expansions calculated the phase diagram for the polymer chain in 
both $2D$ and $3D$ (Euclidean) space. These phase diagrams show the existence of 
the surface attached globule ({\bf SAG}) phase in qualitative agreement with the 
one found earlier for the gasket. It is important to note that in the thermodynamic 
limit, the {\bf SAG} phase has the same free energy per monomer as the {\bf DC} 
phase, and the transition between them is a surface transition.

\begin{figure}[t]
\begin{center}
\includegraphics[width=8cm]{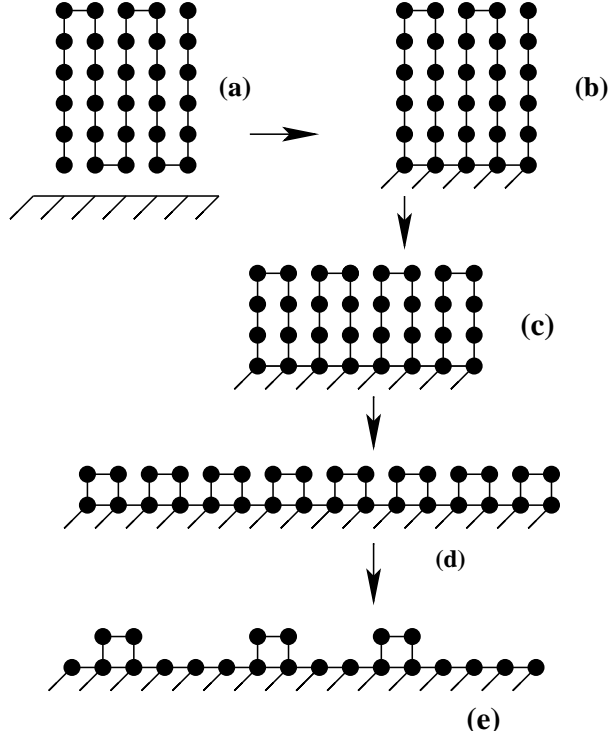}
\end{center}
\caption{ Schematic diagrams of self-avoiding walk conformation at T=0. 
Below critical value $\omega_{c1}$,(a) the number of monomers at the 
surface are zero and polymer is in the desorbed state; (b) the structure 
shown in Fig.(a) gets adsorbed on the surface at critical value $\omega_{c1}$. 
Average number of pair of neighbouring monomers on the surface $<N_c>$ scales 
as $N^{d-1}$; (c) $<N_C>$ increases with $\omega$; (d) At $\omega_{c2}$, $<N_c>$
is maximum; (e) $<N_c>$ decreases sharply with further increase in $\omega$.}
\label{fig1}
\end{figure}

As shown earlier \cite{1}, it is easy to understand the {\bf SAG} and its 
boundary in the limit $T=0$.  At $T=0$, the polymer configurations may be 
mapped by Hamiltonian walk as shown in Fig 1(a).  Its bulk energy is the
same as in the {\bf DC} phase and the surface energy is $(\eu - \es) L_{\|}^{d-1} 
+ (d-1) \eu N/L_{\|}$. Minimizing the surface energy with respect to $L_{\|}$, 
we obtain
\be
\mbox{E}_{\mbox{SAG}}= -(d-1) \eu N+ d \eu^{\frac{d-1}{d}}
(\eu - \es)^{\frac{1}{d}} N^{\frac{d-1}{d}}.
\ee
While for the {\bf DC} phase,
\be
\mbox{E}_{\mbox{DC}}= -(d-1) \eu N+ d \eu N^{\frac{d-1}{d}}.
\ee
In the {\bf AC} phase, we have $L_{\bot}=1$, $L_{\|}=N^{1/(d-1)}$, and
the free energy at $T=0$ is
\be
\mbox{E}_{\mbox{AC}} = -(d-2) \eu N -\es N+ (d-1) \eu
N^{\frac{d-2}{d-1}}.
\ee    

Comparing the energies of these phases, we see that {\bf SAG} phase
has lower free energy than the {\bf DC} or the {\bf AC} phases for
$0 \le \es \le \eu$. Thus the lower and upper boundaries of the {\bf SAG}
phase (lines $\w_{c1}$ and $\w_{c2}$ in Fig.2) tend to $\w_{c1} = 1$ and
$w_{c2} = u$ for large $u$.
   
In the case of partially directed polymer in $2D$ we have found \cite{1} exact 
phase diagram of the {\bf SAG} phase. In this case the polymer has different 
behaviour depending on whether it is near the wall perpendicular to the 
preferred direction ({\bf SAG1}) or the wall is parallel to the preferred direction
({\bf SAG2}). The phase boundaries of {\bf SAG1} and {\bf SAG2} have been found
by calculating this orientation dependent surface energy.

Since the analytical approach are limited to very few cases such as fractal 
lattices or the directed walks on $2D$ lattices, we have to resort to numerical
methods to calculate the phase diagram in other cases. A lattice model using
extrapolation of exact series expansions (herein after referred to exact
enumeration method) has been found to give satisfactory results as it takes
into account  the corrections to scaling. To achieve the same accuracy by the 
Monte Carlo method, a chain of about two order magnitude larger than in the
exact enumeration method has to be considered \cite{7}.

In this article we report our results which we have found using exact 
enumeration method both for $2D$ and $3D$ cases.
Let $C_N (N_s, N_u)$ be the number of SAWs of $N$ sites having
$N_s$ monomers on the surface  and $N_u$ nearest neighbour monomer pairs.
We analyzed  the series for $C_N (N_s, N_u)$ up to $N=30$ for square 
lattice and $N=20$ for the cubic lattice; thus extending the series by two
more terms in both cases. 
 
To obtain better estimates of critical points  as well as the phase
boundaries, we extrapolate for large $N$ using ratio method \cite{8}. Let
\be
Z_N(\omega, u) = \sum_{N_s, N_u} C_N(N_s, N_u) \omega^{N_s} u^{N_u},
\ee
be the partition function. Then, the reduced free energy per monomer 
can be written as \cite{8}
\be
G(\omega, u) = \lim_{N\rightarrow\infty} \frac{1}{N} \log Z_N 
(\omega, u).
\ee

The phase boundaries are then found from the maxima of $\frac{\partial^2
G(\omega,u)} {\partial \es^2}$ ($=\frac{\partial \langle N_s
\rangle}{\partial \es}$) and $\frac{\partial^2 G(\omega,u)}{ \partial
\eu^2}$ ($= \frac{\partial \langle N_u \rangle}{\partial \eu}$).

\begin{figure}[t]
\begin{center}
\includegraphics[width=12cm]{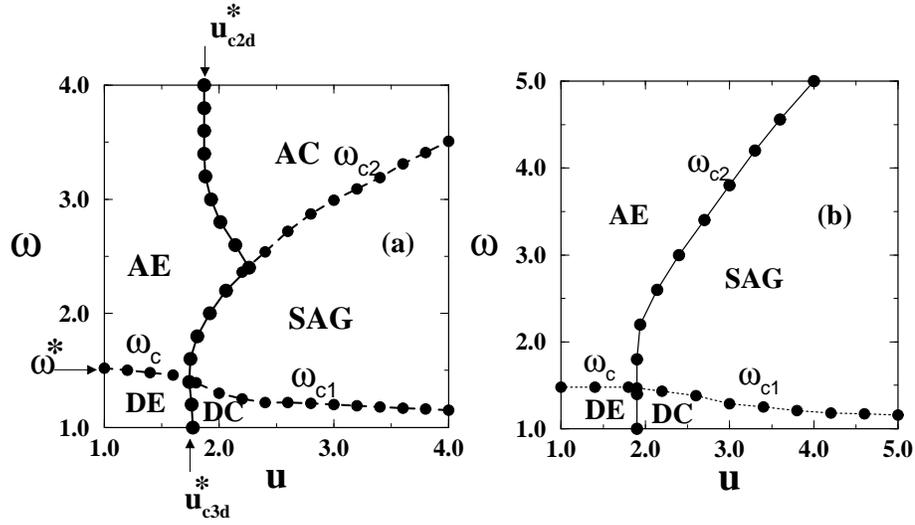}
\end{center}
\caption{ This figure shows the phase diagram of a surface interacting
linear polymer in $3D$ and $2D$ space respectively. The transition lines
were obtained using method of exact-enumeration.}
\label{fig2}
\end{figure}

The phase diagram thus obtained is shown in Fig.2(a) for three dimensions. 
It has five phases which are denoted as desorbed expanded {\bf (DE)}, 
desorbed collapsed {\bf (DC)}, adsorbed expanded {\bf (AE)}, adsorbed 
collapsed {\bf (AC)} and surface attached globule {\bf (SAG)}. Since in 
case of two dimensions, surface is a line, therefore, we do not find adsorbed 
collapsed phase [Fig2(b)]. The value of $u_c$ at $\omega=1$ is found to be 
$1.93 \pm 0.02$ and $1.76 \pm 0.02$ for $2D$ and $3D$, respectively. 
The value of $\omega_c$ at $u=1$ is found to be $2.05 \pm 0.01$ 
and $1.48 \pm 0.02$ for $2D$ and $3D$, respectively. The special adsorption 
line $\omega_c$ separates the {\bf AE} phase from that of {\bf DE} 
phase. The $u_c$ line meets $\omega_c$ line which is a multi-critical 
point ($u_c,\omega_c) = (1.93,1.46)$ for $2D$ and (1.76, 1.38) for $3D$, 
respectively. Values of $3D$ are same as those reported by us earlier \cite{1}.
The value of critical exponents both at the ordinary 
and multi-critical points are in very good agreement with the known 
results \cite{9-10}.

In order to show how the monomers distribute when the chain is in the 
different regimes of the phase diagram, we calculate the value of each 
term in Eq.(4) for given values of $u$ and $\omega$ and plot the results 
for walk up to 20 steps (3D case only) in Fig.(3-6). Here we have plotted 
$\xi/\xi_m$ for eight values of $\omega$, where $\xi = C_N(N_s,N_u) 
\omega^{N_s} u^{N_u}$ ($\xi_m$ being the maximum value of $\xi$) as a 
function of $N_s$ and $N_u$. 
Since the term which contribute most to the sum of Eq.(4) represents
the most probable configurations of the chains, we can infer, from these 
figures about the manner in which the  monomers are distributed in 
different states. 
For example, Fig.(3(a-d) and 4(a-d)) show how the distribution of monomers 
changes as $\omega$ is varied at fixed $u=1.0$ and $1.5$ (corresponding to 
expanded state). It is clear from these figures (3(a) and 4(a)) that $N_s$ 
and $N_u$ are nearly zero around $\omega=1$. In Fig.(3(d)-4(d)), at 
$\omega=3.5$, $N_s$ is around 20 corresponding to the adsorbed expanded state 
of chain. For $u=2.0$ and $2.5$ (correspoding to the collapsed state), 
distribution is shown in Fig.(5(a-d) and 6(a-d). 
For $\omega=1$, the polymer chain is in the bulk as there is any monomer 
on the surface ($N_s \sim 0$) but $N_u$ is large which indicate that
monomers occupy the neighbouring sites. When $\omega$ is chosen equal 
to 2.0 a value which lies in between $\omega_c1$  and $\omega_c2$ line, as 
shown in Fig.2, a fraction of monomers gets attached to the surface 
($N_s \sim 12$) indicating the breaking of translational invariance of the 
globule corresponding to the {\bf SAG} phase. With further increase in 
$\omega$ polymer gets adsorbed in collapsed state where $N_s \sim 20$ and 
$N_u$ is around 8.

\begin{figure}[t]
\begin{center}
\includegraphics[width=14cm]{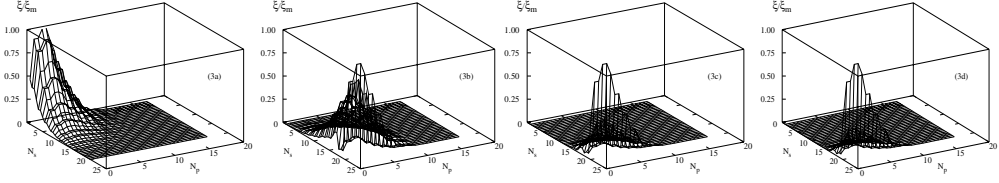}
\end{center}
\caption{Plot of $\xi/\xi_m$ as a function of $N_s$ and
$N_p$. Here $\xi = C_N (N_s, N_p) \omega^{N_s} u^{N_p}$ and $\xi_m$
is the maximum value of $\xi$ for a given value of $\omega$ and $u$.
{\bf (a)} $u = 1.0$, $\omega = 1.0$; {\bf (b)} $u = 1.0$,
$\omega = 2.0$; {\bf (c)} $u = 1.0$, $\omega = 3.0$;
{\bf (d)} $u = 1.0$, $\omega = 3.5$} 
\label{fig3}
\end{figure}

\begin{figure}[t]
\begin{center}
\includegraphics[width=14cm]{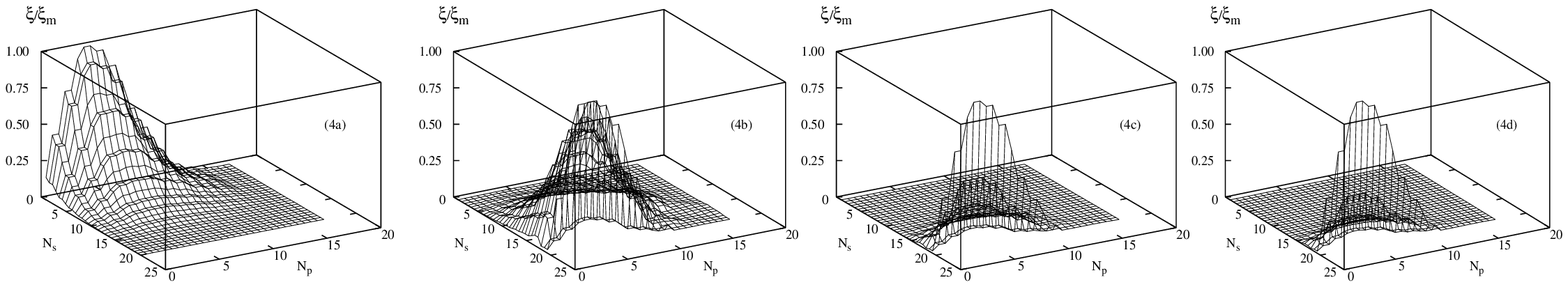}
\end{center}
\caption{Same as Fig.3, {\bf (a)} $u = 1.5$, $\omega = 1.0$; 
{\bf (b)} $u = 1.5$, $\omega = 2.0$; {\bf (c)} $u = 1.5$, $\omega = 3.0$;
{\bf (d)} $u = 1.5$, $\omega = 3.5$} 
\label{fig4}
\end{figure}

\begin{figure}[t]
\begin{center}
\includegraphics[width=14cm]{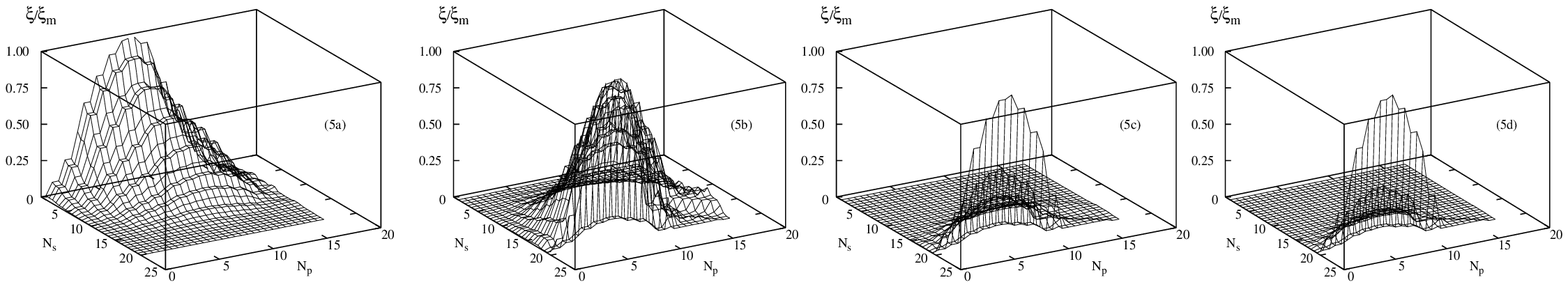}
\end{center}
\caption{Same as FIG.3, {\bf (a)} $u = 2.0$, $\omega = 1.0$; 
{\bf (b)} $u = 2.0$, $\omega = 2.0$; {\bf (c)} $u = 2.0$, 
$\omega = 3.0$; {\bf (d)} $u = 2.0$, $\omega = 3.5$}.
\label{fig5}
\end{figure}

\begin{figure}[t]
\begin{center}
\includegraphics[width=14cm]{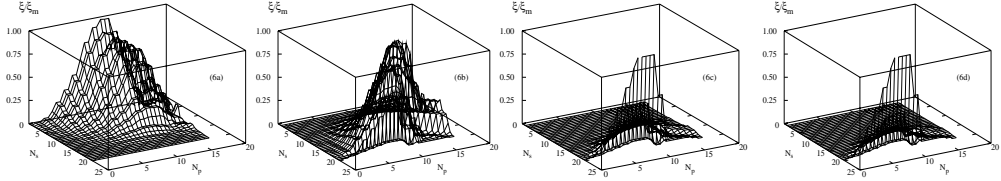}
\end{center}
\caption{Same as FIG.3, {\bf (a)} $u = 2.5$, $\omega = 1.0$; 
{\bf (b)} $u = 2.5$, $\omega = 2.0$; {\bf (c)} $u = 2.5$, 
$\omega = 3.0$; {\bf (d)} $u = 2.5$, $\omega = 3.5$}.
\label{fig6}
\end{figure}

We also calculate number of neighbouring pairs of monomers at the surface 
{\it i.e.} $<N_c>$.  This can be calculated from the expression

\begin{figure}[t]
\begin{center}
\includegraphics[width=10cm]{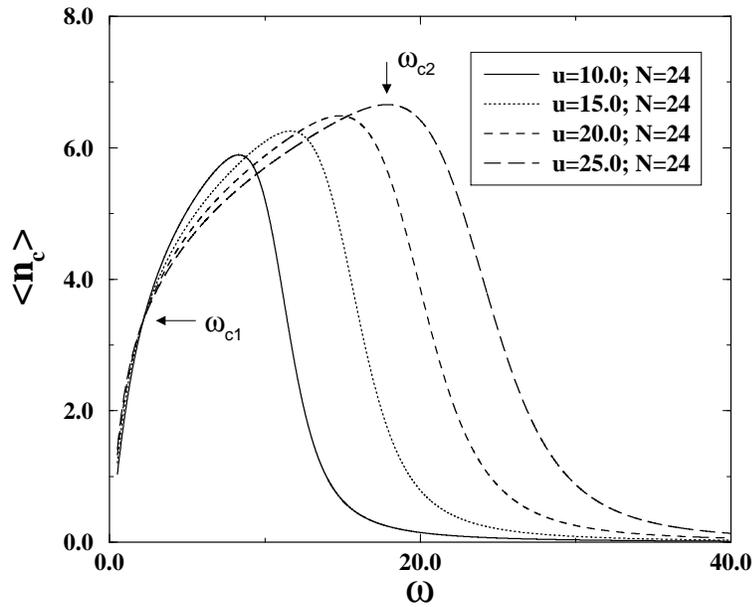}
\end{center}
\caption{Diagram shows the variation of number of monomer pairs on 
the surface $<N_c>$ with $\omega $.}
\label{fig7}
\end{figure}

\be 
<N_c>= \frac{\sum N_c Z_N(\omega,u)}{\sum  Z_N(\omega,u)}
\ee 
 
The variation of $<N_c>$ with $\omega$ for different value of 
$u=10,15,20,25$ is shown in Fig.7 for two dimensions. At such a high 
value of $u$ the most likely structure is of Hamiltonian walk as shown 
in Fig.1(a). Below certain value of $\omega_{c1}$, this structure remains 
in bulk and $<N_c>$ and $<N_s>$ are almost zero.  At certain critical 
value  $\omega_{c1}$, this structure gets stuck on the surface (Fig.1(b)).  
At this value of $\omega$, the $<N_c>$ is equal to $<N_s>$ and goes as 
$N^{d-1}$. As $\omega$ increases $<N_c>$ also increases and reaches to its 
maximum at $\omega_{c2}$. The corresponding configuration is shown in Fig.1(d). 
With further increase in $\omega$, polymer gets adsorbed on the surface 
with decrease in $<N_c>$ as shown in Fig.7 and Fig.1(e).

The results given above should provide ample evidences in favour of the 
existence of the {\bf SAG} phase. The phase boundaries $\omega_{c1}$ and 
$\omega_{c2}$ tend to 1 and $u$ respectively as $T$ goes to zero and 
is independent of dimension. We once again emphasized that the transition 
associated with $\omega_{c1}$ line is surface transition.

We thank Deepak Dhar and R. Rajesh for many helpful discussions. We would 
also like to thank the organizers for giving us opportunity to present 
this work. Financial assistance from $INSA$, New Delhi and DST, New Delhi 
are acknowledged.

yashankit@yahoo.com  \\ 
giri@cts.iitkgp.ernet.in \\
ysingh@banaras.ernet.in

\end{document}